\documentclass[aps,prl,reprint]{revtex4-1}
\usepackage[usenames,dvipsnames]{color}
\usepackage{amsmath,array,amssymb}
\usepackage{graphicx}
\usepackage{hyperref}
\usepackage{textcomp}
\usepackage{braket}
\usepackage{multirow}
\usepackage{indentfirst}
\usepackage[FIGTOPCAP,raggedright,nooneline,bf,footnotesize]{subfigure}
\usepackage{paralist}
\usepackage{overpic}

\renewcommand{\eqref}[1]{Eq.~(\ref{#1})}

\DeclareMathOperator{\sech}{sech}

\begin{document}
\title{Modification of phonon processes in nanostructured rare-earth-ion-doped crystals}
\author{Thomas Lutz$^{1}$, Lucile Veissier$^{1}$, Charles W. Thiel$^{2}$, Rufus L. Cone$^{2}$, Paul E. Barclay$^{1}$ and Wolfgang Tittel$^{1}$}
\address{
$^1$Institute for Quantum Science and Technology, and Department of Physics \& Astronomy, University of Calgary, Calgary Alberta T2N 1N4, Canada
\\
$^2$Department of Physics, Montana State University, Bozeman, MT 59717 USA\\
$^*$Corresponding author: thomasl@ucalgary.ca
}

\begin{abstract}
Nano-structuring impurity-doped crystals affects the phonon density of states and thereby modifies the atomic dynamics induced by interaction with phonons. We propose the use of nano-structured materials in the form of powders or phononic bandgap crystals to enable or improve persistent spectral hole-burning and coherence for inhomogeneously broadened absorption lines in rare-earth-ion-doped crystals. This is crucial for applications such as ultra-precise radio-frequency spectrum analyzers and optical quantum memories. As an example, we discuss how phonon engineering can enable spectral hole burning in erbium-doped materials operating in the convenient telecommunication band, and present simulations for density of states of nano-sized powders and phononic crystals for the case of Y$_2$SiO$_5$, a widely-used material in current quantum memory research. 
\end{abstract}
%


\maketitle
\newpage
Rare-earth-ion-doped crystals have been studied for decades because of their unique spectroscopic properties \cite{macfarlane1987,Liu2005,Sun2002} that arise since their 4$f$-electrons do not participate in chemical bonding. At cryogenic temperatures they can offer narrow linewidths \cite{Bottger2006a,Thiel2011, Thiel2014a} together with the possibility to spectrally tailor their broad inhomogeneous absorption lines \cite{Liu2005}. These properties have led to many applications, including optical quantum memories \cite{Lvovsky2009,Sangouard2011,Bussieres2013},  signal processing \cite{Babbitt2014,Saglamyurek2014}, laser stabilization \cite{Sellin1999,Thorpe2011,Thiel2014b}, as well as ultra-precise radio frequency spectrum analyzers \cite{Lavielle2003,Schlottau2005}. Quantum memory implementations in rare-earth-ion-doped (REI-doped) crystals, such as those based on electromagnetically-induced transparency \cite{fleischhauer_dark-state_2000}), atomic frequency comb \cite{de_riedmatten_solid-state_2008}, or controlled reversible inhomogeneous broadening \cite{kraus_quantum_2006}, crucially rely on long coherence and spin-state lifetimes to achieve high efficiency and long storage time.\\ 

Operating REI-doped crystals at low temperatures generally improves material properties. Yet, even at temperatures below 2 K, spin-lattice relaxation, i.e. thermalization of spins via interaction with phonons, still restricts lifetimes of spin states, reducing the ability to spectrally tailor the material. Furthermore, by contributing to spectral diffusion \cite{Bottger2006,Sinclair2010} and two-phonon elastic scattering processes \cite{Sun2012}, lattice vibrations limit coherence times. We propose the use of nano-structured materials to overcome these limitations. This is achieved by tailoring the phonon density of states \cite{liu_chapter_2007} to restrict phonon processes, an approach not limited to REI-doped materials but also applicable to other impurity-doped materials such as color centers in diamond \cite{jahnke_electronphonon_2015}. The result is an improved performance for all applications based on spectral hole burning, or that require long coherence times, by providing long-lived spin states while simultaneously suppressing phonon-driven decoherence. We note that the modification of population dynamics between REI crystal field levels in small powders, possibly related to phonon restriction, has been observed in the context of understanding luminescence dynamics \cite{Yang1999,Liu2002} - but not yet to improve spectral hole burning or optical coherence properties. \\ 

In this paper, we propose a general procedure to identify the frequency range of detrimental lattice vibrations and restrict the relevant phonon processes in  impurity-doped solids. As an example, we apply this approach to erbium-doped materials. First, we recall the mechanisms causing detrimental electronic spin flips and discuss how the suppression of the direct phonon process influences spin-state lifetimes and spectral diffusion. Then, two methods for nano-structuring REI-doped crystals to suppress the direct phonon process affecting electronic spin states in Er$^{3+}$:Y$_2$SiO$_5$ are analyzed: powder materials, which introduce a cutoff frequency in the phonon density of states, and phononic band gap crystals with a structure optimized for the targeted material and application. Numerical simulations for Er$^{3+}$:Y$_2$SiO$_5$ are presented.\\

Relaxation of optically pumped spin populations back to thermal equilibrium by spin flips is caused by phonons primarily via three different processes \cite{Orbach1961,Kurkin1980}: the direct process where phonons resonant with the spin transition induce a spin flip, the Orbach process where two resonant phonons produce a spin flip via excitation to, and relaxation from, excited crystal field levels, and the off-resonant Raman process where two phonons lead to a spin flip via inelastic scattering. Levels involved in spectral hole burning, typically hyperfine or Zeeman levels, are mostly separated by MHz to GHz. The rates of phonon processes are proportional to the phonon density of states in the material as discussed in \cite{Orbach1961}. Interestingly, the Debye approximation for the density of states 
\begin{equation}
\rho(\omega)\propto V\omega^2
\label{eq:debye}
\end{equation}
with V the volume of the particle and $\omega$ the frequency, breaks down for small structures. In this regime, the density of states becomes discrete \cite{Gschneider2007} and acquires a cutoff frequency below which no vibrational modes are possible and thus no phonons can exist \cite{liu_chapter_2007}. We predict that an ensemble of sufficiently small REI-doped crystals, or an appropriate REI phononic band gap material, will allow suppression of the direct phonon process responsible for relaxation between the spin states. \\
\begin{figure}[t]
\centering
\includegraphics[width=0.9\columnwidth]{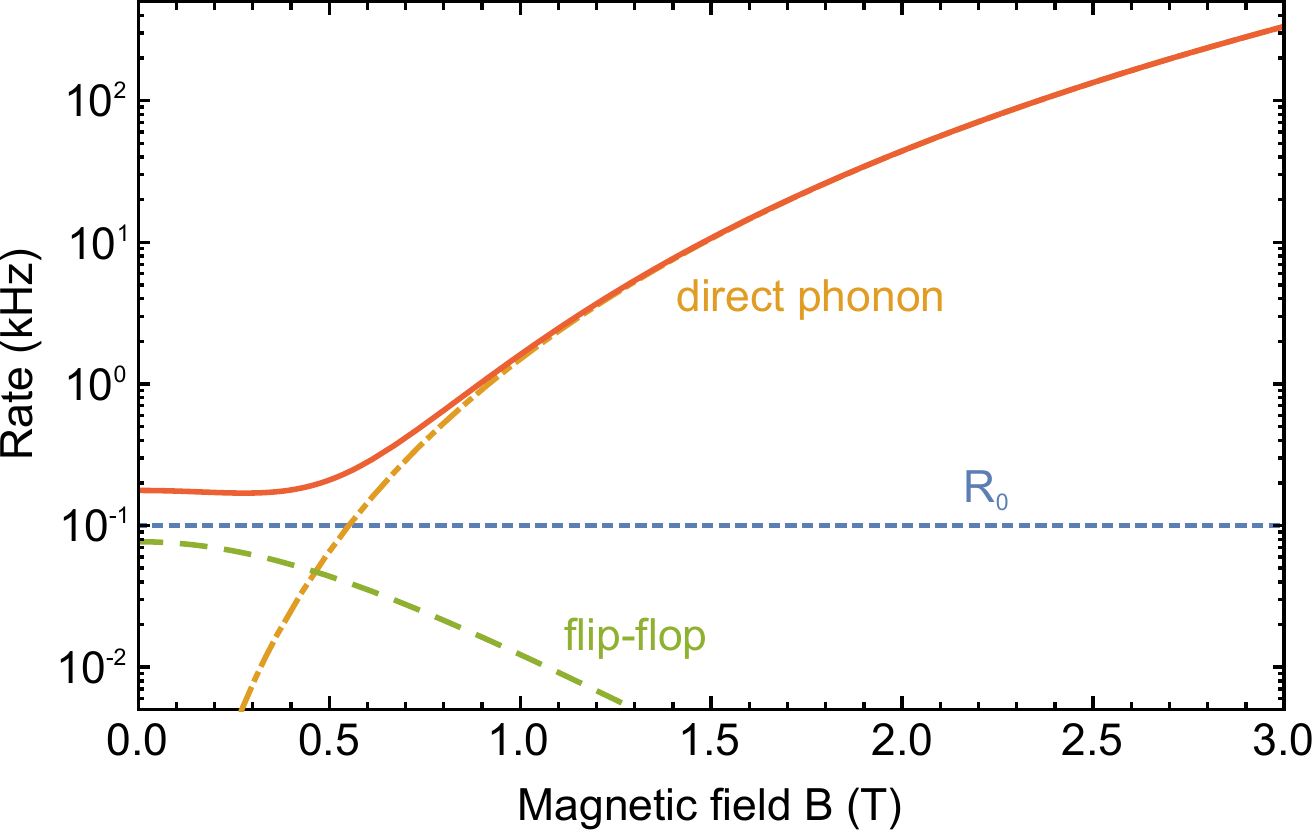}
\caption{Magnetic field dependence of the spin-flip rate. As described by Equation \ref{eq:rates}, the total rate $R$ (solid red line) is composed of a constant term $R_0$ (dotted blue line), a spin flip-flop term (green dashed line) and a direct phonon term (orange dash-dotted line), dominant above 1 T. We assume $T=3$ K, $g=14$, $R_0 = 0.1$ kHz, $\alpha_{\rm D} = 5 \times 10^{-4}$ kHz/T$^5$ and $\alpha_{\rm ff} = 2$ kHz.}
\label{fig:eryso_rate}
\end{figure}

As an example, we consider extending lifetimes of the ground-state Zeeman levels for persistent spectral hole burning and reducing spectral diffusion in erbium-doped crystals, particularly appealing for quantum memory applications due to the $^{4}I_{15/2} \leftrightarrow ^{4}I_{13/2}$ transition of Er$^{3+}$ ions around 1530 nm and long coherence times. Three important cases are Er$^{3+}$:Y$_2$SiO$_5$, for which optical coherence lifetimes of up to 4.4 ms have been observed \cite{Bottger2009}, Er$^{3+}$:LiNbO$_3$, exhibiting a large oscillator strength accompanied by a very broad inhomogenous linewidth \cite{thiel_optical_2010}, and Er$^{3+}$:KTiOPO$_4$, which has shown narrow homogeneous linewidths down to 1.6 kHz \cite{bottger_decoherence_2015}. All Er-doped crystals so far eluded efficient quantum-state storage \cite{Hastings-Simon2008,Lauritzen2008,Lauritzen2011,thiel_optical_2010} since no efficient persistent hole burning involving electronic Zeeman levels \cite{Macfarlane1987b} has been achieved. The only material that has shown long spin-state lifetimes is an Er-doped glass fiber \cite{er-fiber_spectro}, in which disorder broadens the spin transition and effectively weakens the coupling between Er$^{3+}$ ions. This reduces the relaxation rate at low magnetic field. Introducing controlled disorder in crystals is potentially an alternative approach to improve spin lifetimes however, disorder can be extremely detrimental for the coherence properties, as it is the case in impurity-doped glasses. Indeed, the latter are restricted in their applications since coherence times are intrinsically limited to a few $\mu$s or less by low frequency disorder mode dynamics that are present in any amorphous material \cite{macfarlane_optical_2006}. The goal of our study is to propose a method to simultaneously achieve long coherence and spin lifetimes, which excludes, e.g., Er-doped fibers - at least given current knowledge.\\

Phonons can limit the lifetime of persistent spectral holes, but other processes can also contribute, for example  paramagnetic interactions of REI with neighbouring ions or with other impurities in the host. 
For erbium-doped materials such as Er$^{3+}$:Y$_2$SiO$_5$, Er$^{3+}$:LiNbO$_3$ or Er$^{3+}$:KTiOPO$_4$, the total spin-flip rate at temperatures below 4 K can be approximated by \cite{Bottger2006,thiel_optical_2010,bottger_decoherence_2015}
\begin{equation}
\begin{aligned}
R(B,T) = R_{0} + \, \alpha_{\rm ff} \, g^4 \sech ^2 \left( \frac{g \mu_{B} B} {2 k T} \right) \\ 
+ \, \alpha_{\rm D} \, g^{3} B^{5} \coth \left( \frac{g \mu_{B} B} {2 k T} \right).
\end{aligned}
\label{eq:rates}
\end{equation}
The second term represents the average rate of mutual spin flip-flops, i.e. exchange of spin states between two Er ions, with $\alpha_{\rm ff} $ a constant, $g$ the $g$-factor of the ions, $\mu_B$ the Bohr magneton and $k$ the Boltzmann constant. The last term corresponds to the direct phonon process and is proportional to $B^4 T$ for $g \mu_{B} B<2 k T$, where the parameter $\alpha_{\rm D}$ characterizes the strength of the phonon coupling \cite{Orbach1961}. The constant $R_0$ includes all residual relaxation processes. The Raman and Orbach rates are not considered here since they are strongly suppressed at low temperature \cite{Kurkin1980}, particularly when the crystal-field splitting is much larger than $k T$ as is the case for the materials considered here.
The magnetic field dependence of the spin-lattice relaxation is illustrated in Figure \ref{fig:eryso_rate}, which shows the total rate and its different components. Values of the parameters vary for the three materials cited above. The $R_0$ term is negligible in Er:LiNbO$_3$ \cite{thiel_optical_2010}, whereas the spin flip-flop rate is very small for Er:Y$_2$SiO$_5$ \cite{Bottger2006}. Er:KTiOPO$_4$ has unusually small $R_0$ and  weak spin flip-flop process, so that only the direct phonon process significantly contributes to the field-dependent spectral diffusion, even at low field strengths \cite{bottger_decoherence_2015}. However, in all cases the direct phonon process becomes dominant for magnetic field strengths of 1 T and higher.\\

\begin{figure}[]
\centering
\includegraphics[width=0.9\columnwidth]{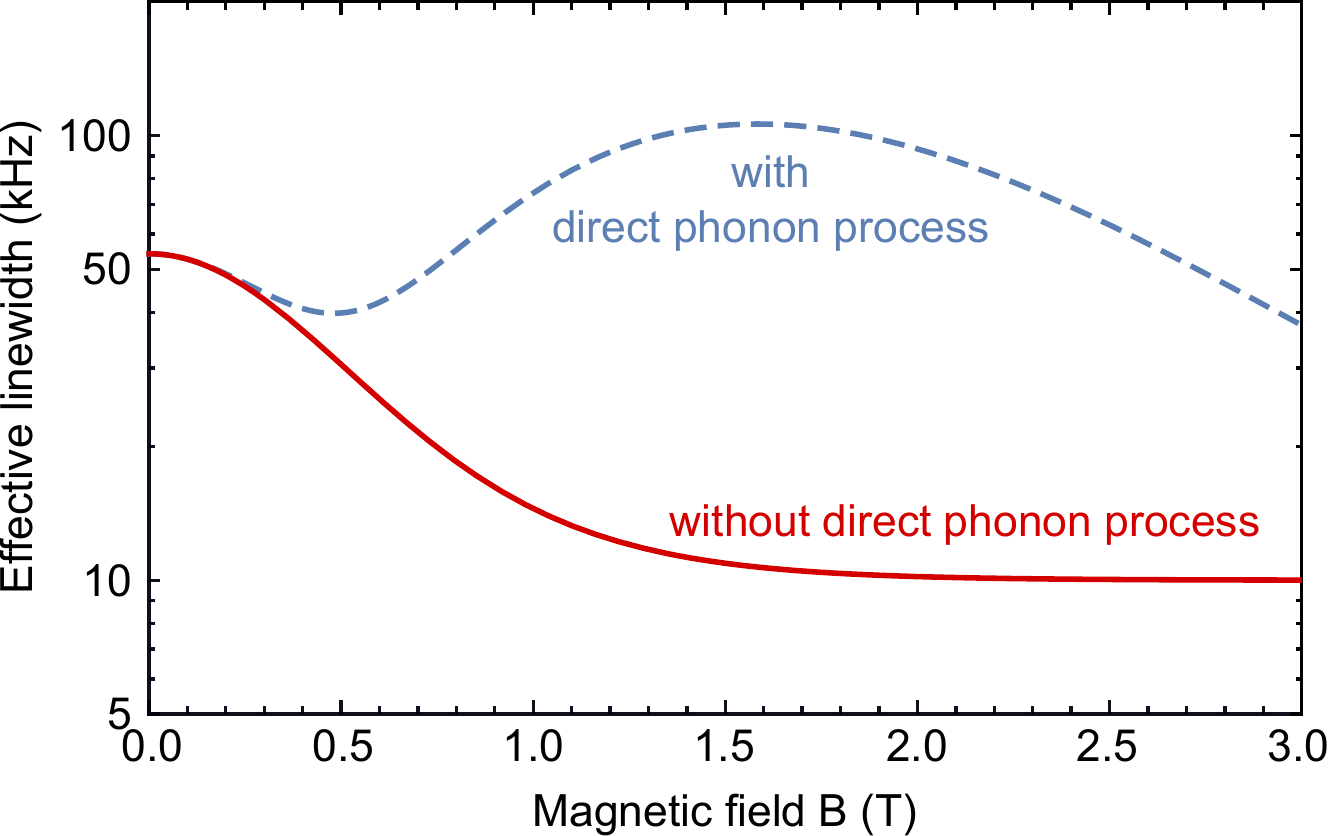}
\caption{Effect of the direct phonon process suppression on the effective homogeneous linewidth. As described by Equation \ref{eq:Gammaeff}, $\Gamma_{\rm eff}$ after $t = 10 \, \mu$s is shown as a function of the magnetic field strength in the presence (blue dashed line) and in the absence (red solid line) of the direct phonon process. We assume the same parameters as in Fig. \ref{fig:eryso_rate} and, in addition, $\Gamma_{\rm max} = 5 \times 10^{10}$ Hz, $\Gamma_0 = 10$ kHz.}
\label{fig:SD}
\end{figure}

From the above discussion, the optimum operating point would seem to be at magnetic fields below 0.5 T; however, to reduce spectral diffusion and improve coherence properties, erbium-ion-doped-crystals often require  magnetic fields of a few Tesla. It has been shown that the maximum linewidth due to spectral diffusion can be described~by~\cite{Bottger2006}
\begin{equation}
\Gamma_{\rm SD}(B,T) =  \Gamma_{\rm max}  \sech ^2 \left( \frac{g \mu_{B} B} {2 k T} \right)  \, ,
\label{eq:GammaSD}
\end{equation}
with $\Gamma_{\rm max} $ the full-width at half-maximum frequency broadening caused by magnetic dipole-dipole interactions between Er ions. One can see that $\Gamma_{\rm SD}$ decreases with magnetic field and spectral diffusion can thus be suppressed at high fields due to the increased magnetic order in the system. This leads to narrow homogeneous linewidths at high fields. However, high fields lead to increased spin-lattice relaxation (see Figure \ref{fig:eryso_rate}). In addition to being detrimental to persistent spectral hole burning, this results in faster broadening of the spectral line to its maximum value, as shown in the effective linewidth after a time delay t \cite{Bottger2006}
\begin{equation}
\Gamma_{\rm eff}(t) = \Gamma_{0} + \, \frac{1}{2} \Gamma_{\rm SD}  \left( 1 - e^{-Rt} \right),
\label{eq:Gammaeff}
\end{equation}
with $\Gamma_0$ the homogeneous linewidth without spectral diffusion. Therefore, removing the direct phonon process will not only improve persistent spectral hole burning, but also inhibit spectral diffusion, especially at magnetic fields around 1 T (see Figure \ref{fig:SD}).\\

We now identify the range of frequencies at which phonons must be suppressed. In our example of erbium-doped materials, the goal is to eliminate phonons below the frequency corresponding to the maximum Zeeman splitting under an external magnetic field of around 1 T. 
In Er:Y$_2$SiO$_5$ the $g$-factor along the $b$-axis is 13.6 \cite{Bottger2002}, leading to a splitting of 190 GHz/T. As we will describe next, inhibiting phonons at such high frequency requires small structures. For other types of paramagnetic centers or nuclear hyperfine states, the energy splitting of the spin states is usually smaller, so that larger structures can inhibit the detrimental phonon processes. 
Furthermore, it is possible to suppress phonons resonant with transitions between crystal-field levels, depending on the energy level splitting. In consequence, two-phonon processes such as Raman or Orbach processes can be suppressed. This would also lead to improved coherence lifetimes for materials in which two-phonon elastic scattering is responsible for dephasing \cite{Sun2012}. More generally, suppressing phonons will enable to increase the range of working temperatures for all impurity-doped materials that have properties limited by phonon processes.\\


We first discuss the case of small crystals. To suppress phonon processes we use the fact that the density of states of a particle features a cutoff at 
\begin{equation}
\nu_{min}= \eta_{min} \frac{c}{\pi d} ,  
\label{eq:cutoff}
\end{equation}
with $c$ the sound velocity, $d$ the diameter of the crystal, and $\eta_{min}$ a numerical constant that is 2.05 for a spherical particle \cite{Lamb1881}. Below this cutoff no phonons are allowed. In the region above the cutoff, the phonon density of states exhibits gaps before it transitions into a continuous function described by Equation \ref{eq:debye}.\\ 

To simulate the phonon density of states for a spherical particle, the eigenmodes must first be calculated. They can be separated into spheroidal-transverse, spheroidal-longitudinal, and torsional modes. The characteristic equations needed to calculate their frequencies have been described in a quantized form by Takagahara \cite{Takagahara1996}. Each eigenmode is identified using three quantum numbers: the radial quantum number $j$, the angular momentum $l$, and $\sigma$, which specifies the type of mode. Using those eigenmodes, the phonon density of states can be written as \cite{Meltzer2000}
\begin{equation}
\rho{(\omega)}=\sum_{j,l,\sigma}(2l+1)\frac{\Delta\omega}{(2\pi)^2}\left[(\omega-\omega_{l,j}^{\sigma})^2+\frac{\Delta \omega}{2}^2\right]^{-1} \, .
\label{eq:pdos}
\end{equation}
Each eigenmode is $2l+1$ times degenerate. We assume that the density of states is composed of Lorentzian shaped lines at the frequency of the respective eigenmode $\omega_{l,j}^{\sigma}$ and linewidth $\Delta\omega$. 
In the work of Meltzer and Hong \cite{Meltzer2000}, the Lorentzian width was assumed to increase with the frequency squared, an empirical assumption that described their particular experimental data adequately. However, since there is no theoretical framework to support that assumption, the width is assumed to be constant here. \\

\begin{figure}[]
\centering
\includegraphics[width=0.9\columnwidth]{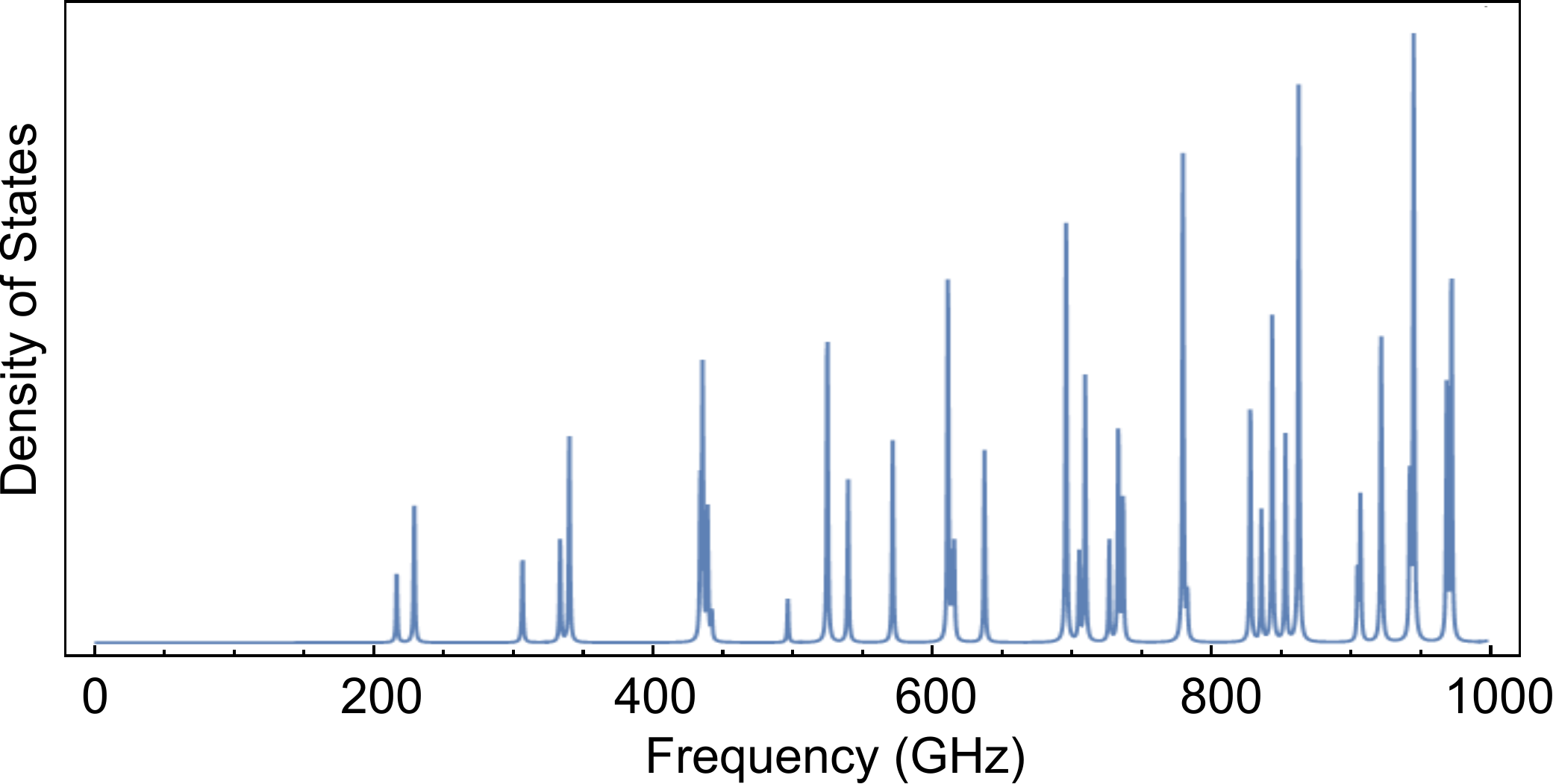}
\caption{Simulated phonon density of states in a Y$_2$SiO$_5$ nanoparticle with a diameter of 12 nm. Each eigenmode of the spherical particle is represented by a Lorentzian with a width of 1 GHz.}
\label{fig:pdos}
\end{figure}

In the case of Er$^{3+}$:Y$_2$SiO$_5$ around 1 T, we estimated from Eq. \ref{eq:cutoff} that nanoparticles of 12 nm diameter are suitable to suppress the direct phonon process. The phonon density of states for a Y$_2$SiO$_5$ nanoparticle of 12 nm diameter, calculated according to Equation \ref{eq:pdos}, is shown in Figure \ref{fig:pdos}. Phonon processes at frequencies below 200 GHz are fully restricted. This assumes that there is no phonon propagation between the particles. Simulating the effect of touching particles on phonon propagation is not trivial and subject further theoretical and experimental research. However, for a lightly packed powder immersed in cold helium gas we expect properties approaching those of single, isolated particles. Note that particles separated by electrostatic interaction can be obtained by coating the powder with the appropriate surfactant or by specific syntheses \cite{Park2004}. 
\\

In powders, averaging of anisotropic properties over the random distribution of crystallite orientations generally causes issues in applications such as quantum memories and spectroscopy. However, if the material has a dielectric or magnetic anisotropy, the crystals can be preferentially oriented using external fields \cite{Schuetze2010,Akiyama2010}. This allows the control of parameters, such as Zeeman splittings and the direction of the ion's transition dipole moment. Scattering of laser light by powder samples can be reduced or even eliminated if the size of the crystals is smaller than the wavelength of the interacting light, since the powder can be treated as a bulk material with an effective (average) refractive index.  \\

Crystal powders can be produced using several approaches and spectroscopic properties comparable to bulk crystals have been observed in some cases \cite{perrot_narrow_2013}. %
A thorough review of different growth methods can be found in \cite{Tissue1998} and references therein. Growing nano/microcrystals has the advantage of being relatively inexpensive. An alternative way to produce nanocrystals is to grind a bulk crystal into powder using techniques such as ball milling. With this approach, sizes on the order of 10 nm are achievable by using very small grinding balls \cite{Bork1998,Song2013}. However, reducing dimensions below around 10 nm may result in additional (detrimental) decoherence \cite{Hartog1999,Meltzer2000} and lifetime limiting effects \cite{Aminov1998}.\\

A second possibility to avoid detrimental phonons is to create a phononic band gap material \cite{MonterodeEspinosa1998,Sanchez-Perez1998,Robertson1998} engineered to produce a gap in the density of states around the undesired frequencies. Such structures are employed in the field of optomechanics \cite{Alegre2011,Chan2012}, and are also used for thermal phonon engineering i.e. for thermal metamaterials and thermocrystals \cite{Maldovan2013}. By changing the geometry and overall size of phononic crystal structures, it is possible to optimize the band gaps for a specific application \cite{Bilal2011}.\\

\begin{figure}[t]
\centering
\includegraphics[width=1\columnwidth]{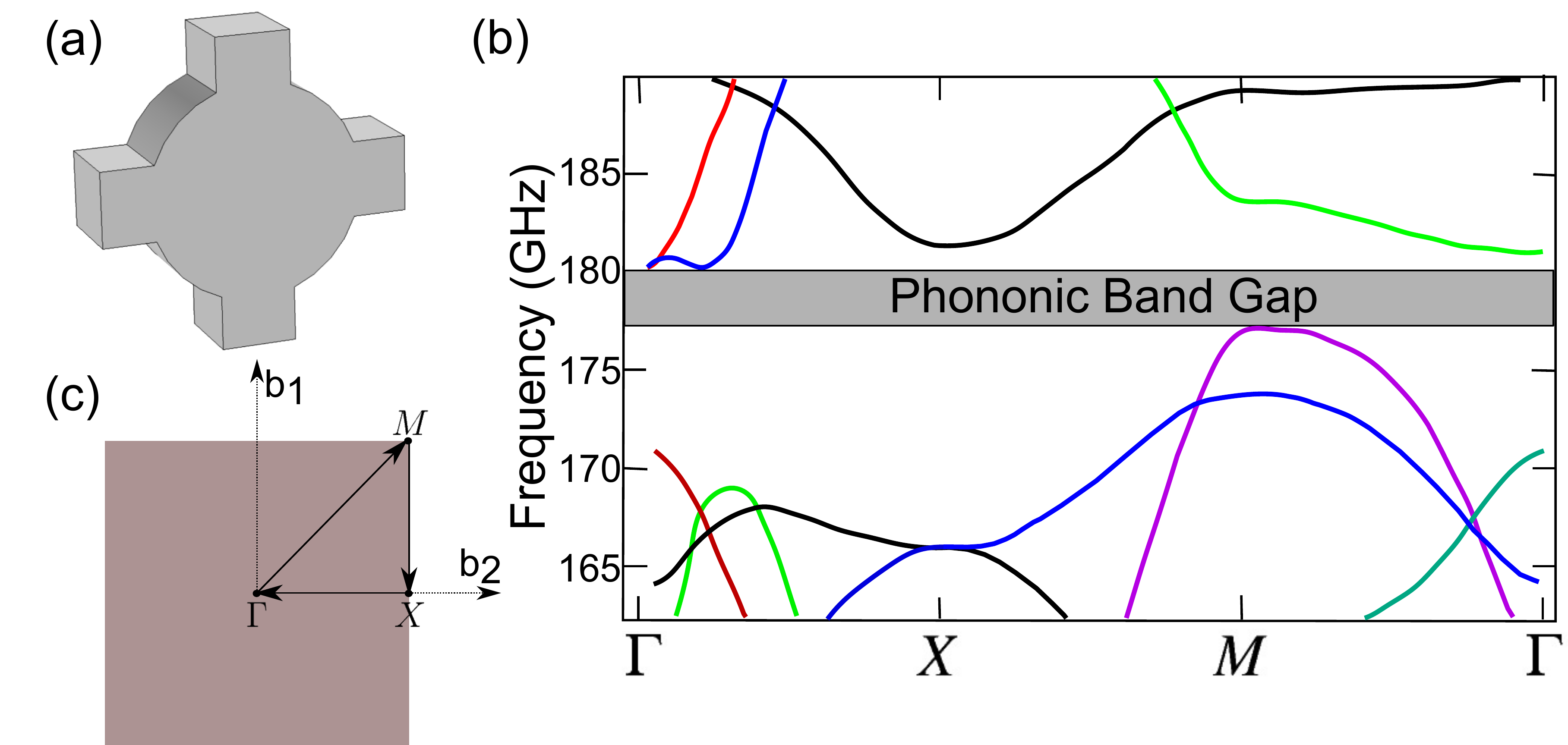}
\caption{(a) Unit cell of the proposed phononic crystal structure. The  radius of the island is 9.5 nm, the width of the arms is 5 nm, and their length 5 nm. This structure is 7 nm thick. (b) Band structure of the phononic crystal in (a). The band gap (shaded grey) around the center frequency of 178 GHz is clearly visible. (c) Reciprocal lattice \cite{Kittel2004} showing the points traversed in the band structure. } 
\label{fig:pnc}
\end{figure}

Phononic crystal materials have the advantage of employing an oriented single crystal. 
In addition, the problem of contact between nanoparticles would be avoided. The fabrication of phononic band gap materials has already been explored for materials such as silicon or gallium arsenide \cite{Petrus2014} and phononic structures with a bandgap at 33 GHz and dimensions around 100 nm have been demonstrated \cite{Goettler2011}. To create phononic crystal structures in currently-used host materials such as Y$_2$SiO$_5$, Y$_3$Al$_5$O$_{12}$ or LiNbO$_3$, thin films of those materials would be a prerequisite. Thin films of LiNbO$_3$ are commercially available and there have been demonstrations of thin single-crystal film fabrication for other REI host materials \cite{Bai1994,Gaathon2012}. Finally, structuring these materials with nanometer precision has been shown to be feasible using focused ion beam milling \cite{Lacour20051421,zhong_nanophotonic_2015}.\\ 

To illustrate this approach for our example of Er:YSiO, we designed and optimized a phononic structure using finite element simulations (COMSOL). The structure is inspired by designs of Safavi-Naeini et al.\cite{Safavi-Naeini2010}. We engineered our structure such that it produces a band gap around 178 GHz with a  width of  3.5 GHz. Figure \ref{fig:pnc} shows the proposed geometry of the phononic crystal unit cell together with its band structure. Even though it is still challenging, the rapid progress in the field of nano-fabrication together with the discovery of helium focused ion beam milling allows structures with these dimensions to be fabricated out of a thin film \cite{Scholder2013,Kalhor2014}. Through additional etching it is possible to remove part of the substrate below the phononic crystal such that the structure suspended and only  connected to the substrate at its borders. This ensures complete phonon suppression.\\

In summary, we propose to restrict phonon processes in REI-doped materials through nano-tailoring in order to improve persistent spectral hole burning and reduce spectral diffusion. We briefly discussed two methods towards this end  -- the use of nano-crystals and engineered phononic band gap structures -- and described how our concept can be applied to well-studied Er-doped crystals. In turn, this would allow realizing optical quantum memories operating at convenient telecommunication wavelengths and increase the performance of  other applications that are based on persistent spectral hole burning or that rely on narrow homogeneous line widths.

\section*{Acknowledgments}
The authors acknowledge support from Alberta Innovates Technology Futures (ATIF), the Natural Sciences and Engineering Council of Canada (NSERC), the Canadian Institute for Advanced Research (CIFAR) of which W.T. is a Senior Fellow, and the National Science Foundation of the United States (NSF) under  Award No. PHY-1212462,  No. PHY-1415628, and No. CHE-1416454.\\

\end{document}